\documentclass[twocolumn,showpacs,preprintnumbers,amsmath,amssymb,showkeys]{revtex4}
\usepackage{graphicx}
\usepackage{dcolumn}
\usepackage{bm}

\begin{document}
\title{Thermally activated charge carriers and mid-infrared optical 
excitations in quarter-filled CDW systems}

\author{I. Kup\v{c}i\'{c}
}                     
\affiliation{Department of Physics, Faculty of Science, University of Zagreb, 
POB 331,  HR-10\,002 Zagreb, 
Croatia}
%
%
\begin{abstract} 
The optical properties of the quarter-filled single-band CDW systems 
have been reexamined in the model with the electron-phonon coupling 
related to the variations of electron site energies.
	It appears that the indirect, electron-mediated coupling
between phase phonons and external electromagnetic fields  
vanishes for symmetry reasons, at variance with the infrared selection 
rules used in the generally accepted microscopic theory.
	It is shown that the phase phonon modes and the electric fields
couple directly, with the coupling constant proportional to the 
magnitude of the charge-density wave.
	The single-particle contributions to the optical
conductivity tensor are determined for the ordered CDW state 
and the related weakly doped metallic state by means of the Bethe--Salpeter 
equations for elementary electron-hole excitations.
	It turns out that this gauge-invariant approach establishes
a clear connection between the effective numbers of residual, 
thermally activated and bound charge carriers.
	Finally, the relation between these numbers and the  activation energy of dc 
conductivity and the optical CDW gap scale is explained in the way consistent
with the conductivity sum rules.
\end{abstract}
\maketitle
\section{Introduction}
%
%

The electrodynamics of low-dimensional multiband systems in which strong
correlations responsible for formation of localized electronic states
coexist with  correlations responsible for charge-density-wave ordering (CDW) 
still attract great attention.
	The vanadium chain compound BaVS$_3$ is a typical example of such 
(magnetic) metals with the commensurate CDW instability 
\cite{Forro00,Fagot,Kezsmarki,Mitrovic}.
	In this system, basic electrodynamic properties are related
to one wide band, while  magnetic anomalies are associated with more or less
localized states in two narrow bands.
	The relations between the mid-infrared (MIR) energy scales, measured 
at temperatures below the critical CDW  temperature, 
the activation energies of transport coefficients 
and the related effective numbers of conduction electrons 
are not entirely clear \cite{Kezsmarki}.
	In order to explain these relations  in simple physical terms, 
one needs, in the first place, a precise description of the CDW 
coherence effects in the quarter-filled wide band.
	This problem that is still not completely solved for the single-band 
case is in the focus of this work.

The previous theoretical  work 
\cite{LRA,Schulz,Rice,Gruner,Fukuyama,Brazovskii,Littlewood,Artemenko,Maki,Brazovskii93,Solyom,Giamarchi,Kim,McKenzie}
on the electrodynamics of the single-band CDW systems 
followed two basic routes.
	First, the collective contributions have been studied to explain 
the low- and zero-temperature properties of the incommensurate CDW systems, 
with the emphasis on the non-linear conductivity regime
\cite{Artemenko,Maki}.  
	The main controversies characterizing early theoretical investigations
are resolved using  accurate symmetry analyses, 
with a particular care devoted to the local field effects
\cite{Brazovskii93}.
	The principal physical problem in these analyses is that 
the coupling between infrared-active collective modes and external electromagnetic
fields is shown as a simple function of the single-particle (interband) 
polarizability \cite{LRA,Rice,Maki,Brazovskii93},
which implies the indirect electron-mediated, rather than the direct, 
photon-phase phonon coupling.
	The high (MIR) frequency analyses \cite{Kim,McKenzie}, on the other hand, were
focussed on the precise description of the single-particle 
excitations, in particular on the rigorous treatment of the square-root
singularity in the optical conductivity spectra at $\hbar \omega = 2 \Delta$.
	It is shown that the controversies related to the high-frequency
optical spectra usually reflect the incorrect treatment of the diamagnetic
current in the transverse response approaches, and, consequently, 
can be easily avoided using the longitudinal response theory \cite{KupcicPB1}.

In this article,  we consider a quarter-filled single-band 
CDW system and a related  tetramerized metallic system    
in which all non-retarded short-range interactions are neglected,
and all coherence effects in the correlation functions associated with the
retarded, phonon-mediated interations
are treated exactly through the known analytical 
form of the current vertices and the inverse effective-mass tensor.
	Using this exactly solvable model, we reexamine several textbook results
\cite{LRA,Schulz,Rice,Gruner} for the dc and optical conductivity of 
commensurate CDW systems.
	The analysis is however limited by the use of several approximations.
	We consider a Holstein-like electron-phonon coupling in which 
the coupling constant is independent of the electron wave vectors.
	Being interested in the optical conductivity spectra obtained 
by the reflectivity measurements, which are characterized by a relatively 
poor resolution in the low-frequency part of the spectra, 
we approximate the total optical conductivity by the sum of 
the collective and  single-particle contributions,
with the local field effects included only in the collective term.
	In the numerical analysis of the single-particle term, we 
consider the simplest case in which two relaxation rates are assumed to be 
independent of frequency and temperature. 
	Finally, the commensurability (Umklapp) effects are present 
in the model  through the shift of the phase phonon frequency to a finite value
and through the commensurate photon-phonon coupling (see Sec.~5).

The article is organized as follows.
	In Sec. 2, the band dispersions of the tetramerized CDW case are 
briefly discussed.
	In Secs.~3 and 4.1, we consider the electron-hole excitations
in the two-band version of the model in which the Umklapp effects are neglected.
	The emphasis is on the clear identification 
of the effectively free and effectively bound charge carriers 
in the underdamped regime
(the interband damping energy is small in 
comparison with the CDW gap), 
their relation to both the activation energy 
of dc conductivity and the MIR scale in the optical conductivity.
	The brief discussion of the optical conductivity spectra
in the overdamped regime, tentatively related to the 
hybridizations/interactions with localized electronic states, 
is given in Sec.~4.2.
	In Sec. 5, the infrared selection rules
for the collective modes in the commensurate CDW systems are 
discussed in some detail.
	It is shown that the (interband) electron-mediated photon-phase phonon
coupling vanishes for symmetry reasons, and that the direct coupling 
is proportional to the magnitude of the charge-density wave.

\section{Tetramerized CDW case}
%
%

%
\subsection{Bloch energies}
%

The starting point of the present analysis is the Q1D tetramerized model in
which the electron-phonon coupling is formulated in terms of 
the displacement vector
\begin{equation}
{\bf u}_i = {\bf u}_{\varphi i} + {\bf u}_{A i}
= \sqrt{\frac{2}{N}} \sum_{{\bf q}\approx 0}
{\rm e}^{{\rm i} {\bf q} \cdot {\bf r}_i} \big[
{\bf u}_{\varphi{\bf q}} \cos \phi_i  -{\bf u}_{A{\bf q}} \sin \phi_i  \big],
\label{eq1}
\end{equation}
%
%
where $\phi_i  = {\bf Q}\cdot {\bf r}_i - \phi$, ${\bf r}_i$ is the position of 
an atom and $\phi$ is  an arbitrary phase.
Two Fourier components of interest, ${\bf u}_{A {\bf q}}$ and 
${\bf u}_{\varphi {\bf q}}$, are functions of two ordinary components 
of the wave vectors ${\bf q} \pm {\bf Q}$, as follows 
${\bf u}_{A {\bf q}}  = (1/\sqrt{2}) \big[{\bf u}_{{\bf q}+{\bf Q}}
+ {\bf u}_{{\bf q}-{\bf Q}} \big]$
and 
${\bf u}_{\varphi {\bf q}}  = ({\rm i}/\sqrt{2}) \big[{\bf u}_{{\bf q}
+{\bf Q}}- {\bf u}_{{\bf q}-{\bf Q}} \big] $.
${\bf u}_{A i}$ and ${\bf u}_{\varphi i}$ correspond 
to two similar lattice deformations which phase is shifted by
$\pi/2$ and which frequency is expected to be strongly renormalized for low
enough temperatures and for the band filling close to the quarter filling
(evidently, ${\bf u}_{\lambda  {\bf q}} 
= {\bf u}^\dagger _{\lambda -{\bf q}}$,
with $\lambda = A, \varphi$ being the phonon branch index).
	${\bf Q} = (0.5 \pi/a, Q_y)$ is the nesting vector of the Fermi surface
of the quarter-filled (CDW) case, or the tetramerization vector in a 
general (metallic) case.
	The related variations of the electron site energies (see Fig.~1), 
which are assumed to be the main mechanism of the electron-phonon 
coupling here, are
\begin{eqnarray}
V({\bf r}_i) &=& V_{\varphi}({\bf r}_i) + V_A({\bf r}_i) 
\nonumber \\
&=& \frac{2g}{\sqrt{N}} \sum_{{\bf q}\approx 0}
{\rm e}^{{\rm i} {\bf q} \cdot {\bf r}_i} 
\big[ u_{\varphi {\bf q}}\sin  \phi_i  + u_{A {\bf q}} \cos  \phi_i \big].
\label{eq2}
\end{eqnarray}
%
%
$g$ is the electron-phonon coupling constant.
	The direct inspection of Fig. 1 shows that the periodic lattice
distortion associated  with the replacement 
$u_{A {\bf q}} \rightarrow \delta_{{\bf q},0} \langle u_{A} \rangle 
+ u_{A {\bf q}}$ leads to the charge transfer of the magnitude $q_{c}$
(determined in Sec.~5) modulated by the wave vector ${\bf Q}$.
	This lattice distortion  makes  ${u}_{A {\bf q}}$ and 
${u}_{\varphi {\bf q}}$ to be the long wavelength phonon modes which are,
respectively, Raman and infrared active.
	The other phonon modes are taken into account later in Sec.~3, and 
are treated on the same footing with the  static disorder.

 \begin{figure}
\begin{center}
\resizebox{0.4\textwidth}{!}{%
  \includegraphics{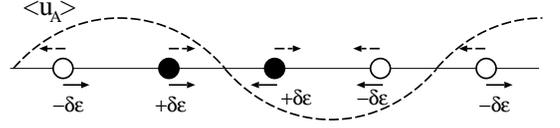}
}
 \caption{
Variations of the electron site energies 
($-\delta \varepsilon$, $+ \delta \varepsilon$, $+\delta \varepsilon$,
$-\delta \varepsilon$) induced by
the periodic (tetramerization) lattice distortion 
$u_{A {\bf q}} \rightarrow \delta_{{\bf q},0} \langle u_{A} \rangle$ and
$\phi = \pi/4$ in Eq.~(\ref{eq1}). 
The solid and dashed arrows illustrate the Raman-active and 
infrared-active ${\bf q}\approx 0$ modes.
The effective charges of the white and the black ions are $+q$ 
and $-q$, respectively.
	Here $\delta \varepsilon = \sqrt{2} \Delta$ and $q=q_c/\sqrt{2}$. 
    }
\end{center}
\end{figure}

The CDW instability of the quarter-filled Q1D electronic systems 
in the case in which the disorder is absent and the electron-phonon coupling 
$g$ is independent of the electron wave vectors is described by the well-known
textbook expressions (see Eqs.~(\ref{eqA1})--(\ref{eqA10}) in Appendix).
	The commensurability effects enter in the theory  through the 
$\phi$ dependent (Umklapp) contributions in  the Bloch energies 
$E_L ({\bf k})$.
	For example, in the 1D limit  ($t_b \rightarrow 0$), 
the four Bloch energies $E_L ({\bf k})$ are 
($L = C, \underline{C}, V,\underline{V}$ is the band index)
\begin{eqnarray}
E_{C,\underline{C}} ({\bf k}) &=& - \sqrt{2t_a^2+ 2\Delta^2 \pm 2t({\bf k})},
\nonumber \\
E_{V,\underline{V}} ({\bf k}) &=&  \sqrt{2t_a^2+ 2\Delta^2 \mp 2t({\bf k})},
\label{eq3} \\ 
t({\bf k}) &=& 
 \sqrt{2\Delta^2 t_a^2 +t_a^4\cos^2 2k_x a+ \Delta^4\cos^2 2\phi},
 \nonumber  
\end{eqnarray}
%
%
with $t_a$ and $t_b$ being the bond energies in the highly conducting 
direction and in the perpendicular direction, respectively,
and $\Delta = g \langle u_{A} \rangle /\sqrt{N}$ is the magnitude of the 
order parameter.	
	When $\Delta$ is small enough, the dispersions of two lowest bands 
can be approximated by 
\begin{eqnarray}
E_{C,\underline{C}} ({\bf k}) &=& 
\tilde{\varepsilon} ({\bf k}) 
\mp  \sqrt{\tilde{\varepsilon}^2  ({\bf k})
- \varepsilon_c ({\bf k})\varepsilon_{\underline{c}} ({\bf k})
- U({\bf k})}. 
\label{eq4}
\end{eqnarray}
%
%
Here $\varepsilon_{c} ({\bf k}) = \varepsilon ({\bf k}) =
-\sum_{\alpha} 2t_{\alpha} \cos k_\alpha a_\alpha$
and $\varepsilon_{\underline{c}} ({\bf k}) = 
|\varepsilon ({\bf k}+{\bf Q})|$ are two
relevant bare dispersions, and
$2 \bar{\varepsilon} ({\bf k}) = \varepsilon_c ({\bf k})
+ \varepsilon_{\underline{c}} ({\bf k})$,
$\tilde{\varepsilon} ({\bf k}) \approx \bar{\varepsilon} ({\bf k})
+ \Delta^2/(2\bar{\varepsilon} ({\bf k}))$ are useful abbreviations.
	The Umklapp processes are represented in the dispersions 
(\ref{eq4}) through
\begin{eqnarray}
U({\bf k}) &\approx&  
\alpha \bigg(1 + \frac{\Delta^2}{4\bar{\varepsilon}^2 ({\bf k})} \bigg)
\frac{2 \Delta^4 - \delta^4({\bf k}) 
- (\delta^*({\bf k}))^4}{4\bar{\varepsilon}^2 ({\bf k})},
\label{eq5}
\end{eqnarray}
%
%
with $\delta({\bf k}) = \Delta \exp \{ {\rm i} \phi ({\bf k}) \}$,
$ \phi ({\bf k}) = {\rm sgn} (k_x) \phi$ and $\alpha = 1$.
	The quarter-filled $\alpha = 0$ limit of Eq.~(\ref{eq4}) 
with $\tilde{\varepsilon}^2  ({\bf k}) \approx 
\bar {\varepsilon}^2  ({\bf k}) + \Delta^2$ corresponds 
to the usual Peierls model which electrodynamic
properties will be studied in more detail in Secs.~3 and 4.

\subsection{Optical conductivity in pure CDW systems}
%
%
In the present context, the main effect of the Umklapp processes
is to shift the renormalized frequency $\omega_{\varphi 0}$ 
of the ${\bf q}=q_x\hat{e}_x=0$ phase phonon mode from zero to a finite value \cite{LRA}.
Together with other pinning mechanisms, this gives  the 
collective contribution to the optical conductivity of the form
(see Fig.~2(a))
\begin{eqnarray}
\sigma_{xx}^{\varphi} (\omega) &=&
\frac{e^2 n}{m^*} \frac{{\rm i} \omega}
{\omega \big(\omega + {\rm i} \gamma_{\varphi {\bf q}}\big) 
- \omega^2_{\varphi {\bf q}}},
\label{eq6}
\end{eqnarray}
%
%
with $m^*$ being the collective mode effective mass.
	In the pure case, where the scattering of electrons on both the 
disorder and the phonon modes is neglected, $\omega_{\varphi {\bf q}}$
and $\gamma_{\varphi {\bf q}}$ in Eq.~(\ref{eq6}) 
are the solutions of the usual functional integral  approach
\cite{Brazovskii}, or the solutions of the Dyson equation shown in Fig.~2(b) 
\cite{LRA,Schulz,Rice}.
	The structure of  $m^*$  and $\omega_{\varphi {\bf q}}$
is discussed in more detail in Sec.~5 and Appendix.

The related single-particle contribution to the optical conductivity 
(Fig.~2(c)) is given, for example, by the gauge-invariant expression
\begin{eqnarray}
\sigma_{\alpha \alpha}^{\rm sp} (\omega) &=&
\frac{{\rm i} \omega}{q^2_\alpha} \chi_{1,1} (q_\alpha, \omega)
= - {\rm i} \omega \alpha (q_\alpha, \omega)
\label{eq7}
\end{eqnarray}
%
%
obtained by means of the longitudinal response theory \cite{KupcicPRB}
(the left-hand side of Fig.~2(c), with ${\bf q} = q_\alpha \hat{e}_\alpha$). 
$\chi_{1,1} (q_\alpha, \omega)$ is the sum of the intra- and interband
charge-charge correlation functions defined in Appendix, and 
$\alpha (q_\alpha, \omega)$ is the related polarizability.


 \begin{figure}[tb]
\begin{center}
\resizebox{0.45\textwidth}{!}{%
  \includegraphics{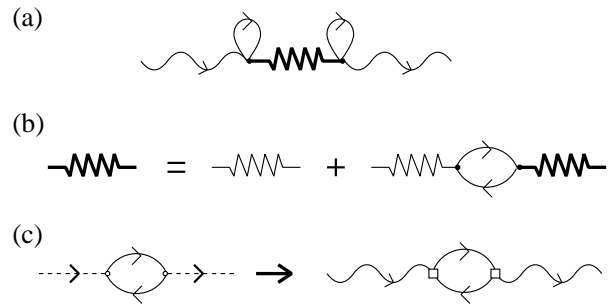}
}
 \caption{
The collective (a) and single-particle (c)  contributions  to the
optical conductivity in pure regime of the present CDW model.
The solid lines are the electron propagators and the zigzag lines 
are the phonon propagators. 
The open circles, the open squares and  the full circles represent
the charge vertices, the current vertices and the
electron-phonon coupling constants, respectively. 
The  dashed line is the external scalar field
$V^{\rm ext} ({\bf q},\omega)$ and the wavy line 
is the related electric field (see Eq.~(\ref{eq13})).  
(b) The  Dyson equation for the phonon Green functions. 
The phonon self-energy is given by the expression (\ref{eqA10}) in Appendix.
}
\end{center}
\end{figure}

\section{Bethe--Salpeter equations}
%
%
We take the collective mode contribution (\ref{eq6}) 
aside and continue with the examination of the (single-particle) 
electrodynamic properties of the $\alpha=0$ two-band model 
of Eqs.~(\ref{eq4}),(\ref{eq5}).
	The Bloch energies are
\begin{eqnarray}
E_{C,\underline{C}} ({\bf k}) &=& \bar{\varepsilon}({\bf k}) 
\mp  \sqrt{(1/4) \varepsilon^2_{\underline{c}c} ({\bf k}) + \Delta^2 }, 
\label{eq8}
\end{eqnarray}
%
with $\varepsilon_{\underline{c}c} ({\bf k}) =
\varepsilon_{\underline{c}} ({\bf k})-\varepsilon_{c} ({\bf k})$.
	To make the analysis more general, we consider two  regimes
of the two-band model close to the quarter filling 
in which an important role is played 
by thermally activated electrons in both transport and optical properties. 
	The first regime corresponds to the ordered  CDW state considered above, 
while the second regime is associated with the metallic state, 
where the doubled Fermi wave vector is $2k_{\rm F} \neq Q_x = 0.5\pi/a$.

\subsection{Relaxation processes on the (quasi)static disorder}
Having in mind the known structure of the  current vertices
related to the bands (\ref{eq8}) \cite{KupcicPB2},
or, alternatively,  calculating the long wavelength charge vertices (\ref{eq29}), 
we can make the generalization of the optical conductivity 
(\ref{eq7}) to the case in which the (quasi)static disorder is present.
	This can be easily done by using the  longitudinal equation-of-motion  
approach \cite{Pines,KupcicPRB}, because this approach treats 
the (intraband) charge continuity equation in a natural way 
(the calculation is  more complicated in the transverse response theory
\cite{KupcicFA}).
	The result (Eqs.~(\ref{eq22}) and (\ref{eq25}))
is equal to Eq.~(\ref{eq7}) with
the adiabatic term $\eta$ in the charge-charge correlation functions 
replaced by two phenomenological damping energies
$\hbar \Gamma_\alpha^{\rm intra}$ and $\hbar \Gamma_\alpha^{\rm inter}$.

The  longitudinal equation-of-motion  approach  is one example of 
the calculations which describe the elementary electron-hole excitations 
in the effective single-particle multiband models in the gauge-invariant
way. 
	The Bethe--Salpeter approach, described below, is another example.
	In both cases, the elementary excitations 
are represented by the Matsubara electron-hole propagator
$$
{\cal D}^{LL'} ({\bf k}, {\bf k}_+, {\bf k}'_+,{\bf k}', \tau)
 =\sum_n {\cal D}^{LL'}_{(2n)} ({\bf k}, {\bf k}_+, {\bf k}'_+,{\bf k}', \tau),
$$
where
\begin{eqnarray}
&& \! \! \! \! \! \! \! \! 
{\cal D}^{LL'}_{(n)} ({\bf k}, {\bf k}_+, {\bf k}'_+,{\bf k}', \tau)
= \frac{(-1)^{n+1}}{\hbar^n n!}
\int_0^{\beta \hbar} \! \!  {\rm d} \tau_1 \ldots 
\int_0^{\beta \hbar} \! \!  {\rm d} \tau_n \,
\nonumber \\
&& \hspace{10mm} \times 
\langle T_{\tau} \big[ H_1'(\tau_1) \ldots H_1'(\tau_n)
c_{L{\bf k} \sigma}^{\dagger} (\tau)c_{L'{\bf k}+{\bf q}  \sigma} (\tau)
\nonumber \\
&& \hspace{15mm} \times
c_{L'{\bf k}' +{\bf q} \sigma}^{\dagger}(0)c_{L{\bf k}' \sigma} (0) 
\big] \rangle 
\label{eq9}
\end{eqnarray}
%
%
and ${\bf k}_+ = {\bf k} + {\bf q}$.
	Here $H_1'$ describes the scattering of conduction 
electrons on the (quasi)static disorder.
	Using this approach, it is possible to  treat the relaxation processes 
on impurities, on lattice imperfections, and even on soft phonons.
	The hybridization with other (uncorrelated) bands can also be 
studied  in this way.

\subsection{Relaxation processes on phonons}

In the presence of  inelastic electron scattering processes 
on phonon modes, or other boson modes in general,
one can use a similar longitudinal diagrammatic approach based on 
the self-consistent Bethe--Salpeter equations for  
$$
\Phi^{LL'} ({\bf k}; {\rm i} \omega_n) \equiv 
\Phi^{LL'} ({\bf k}, {\bf k}_+, {\bf k}'_+,{\bf k}',
{\rm i} \omega_n, {\rm i} \omega_{n+})
$$
(${\rm i} \omega_{n+} = {\rm i} \omega_{n}+ {\rm i} \nu_n$).
	In this case $H_1'$ in Eq.~(\ref{eq9}) represents the 
electron-phonon Hamiltonian (\ref{eqA4}) with the phonon wave vector
${\bf q}'$ running over the entire Brillouin zone and the phonon branch 
index $\lambda$ running over all (elastic and inelastic) scattering channels. 
	$\Phi^{LL'} ({\bf k}; {\rm i} \omega_n)$ is defined by 
the Fourier transform of
${\cal D}^{LL'} ({\bf k}, {\bf k}_+, {\bf k}'_+,{\bf k}', \tau)$,
\begin{eqnarray} 
&& \! \! \! \! \! \! \! \! \! \! \! \! 
{\cal D}^{LL'} ({\bf k}, {\bf k}_+, {\bf k}'_+,{\bf k}', {\rm i} \nu_n)
\\ \nonumber \hspace{15mm}
&& = (\hbar / \beta) \sum_{{\rm i} \omega_n} 
\Phi^{LL'} ({\bf k}, {\bf k}_+,  {\bf k}'_+,{\bf k}',
 {\rm i} \omega_n, {\rm i} \omega_{n+}).
 \label{eq10}
\end{eqnarray}
%
%

 \begin{figure}
\begin{center}
\resizebox{0.45\textwidth}{!}{%
  \includegraphics{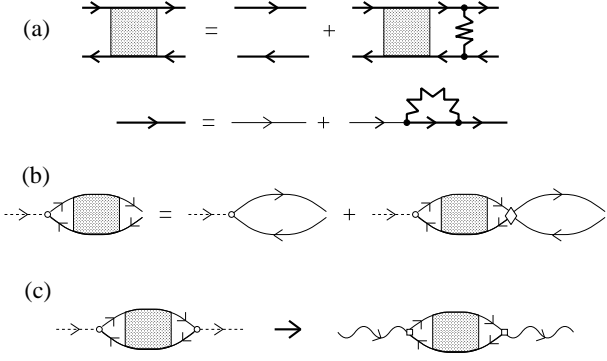}
}
 \caption{(a) The Bethe--Salpeter equations.
(b) The related self-consistent equations for the induced densities 
$\delta n^{LL'} ({\bf k},\omega)$.
The diamond represents the electron-hole self-energy.
(c) The generalization of Fig.~2(c).
    }
\end{center}
\end{figure}

For temperatures and dopings not too close to the critical values 
characterizing the CDW metal-to-insulator phase transition,
the interband electron-phonon coupling in $H_1'$ 
can be neglected (notice that beyond this approximation one needs four band
indices in the electron-hole propagator (\ref{eq9})).
	Now, for 
$G^{LL}_\lambda({\bf k}+ {\bf q}',{\bf k}) \approx 
\sqrt{\hbar/(2M \omega_{\lambda {\bf q}'})}$
$\times gg^{LL}_\lambda({\bf k}+ {\bf q}',{\bf k})$ and 	
$\big| G^{LL}_\lambda({\bf k}+ {\bf q}',{\bf k})\big|^2 \approx 
\big| G_\lambda({\bf q}')\big|^2$,
the Bethe--Salpeter equations \cite{Wolfle} take the form (see Fig.~3(a))
\begin{eqnarray}
&& \! \! \! \!\! \! \! \! \! \! \! \! 
\big[ {\rm i} \hbar \nu_n + E_{LL'}({\bf k},{\bf k}_+) 
+ \Sigma_{L} ({\bf k}, {\rm i} \omega_n ) 
- \Sigma_{L'} ({\bf k}_+, {\rm i} \omega_{n+} ) 
\big]
\nonumber \\
&& \times \Phi^{LL'} ({\bf k}; {\rm i} \omega_n) =
\frac{1}{\hbar} \big[ {\cal G}_{L} ({\bf k}, {\rm i} \omega_n ) 
- {\cal G}_{L'} ({\bf k}_+, {\rm i} \omega_{n+} ) \big]
\nonumber \\
&&  \hspace{10mm}
\times \bigg\{ \delta_{{\bf k}, {\bf k}'} - \frac{1}{\beta \hbar} 
\sum_{\lambda {\bf q}' {\rm i} \nu_m}
\frac{\big| G_\lambda ({\bf q}')\big|^2}{N} {\cal D}_\lambda ({\bf q}', {\rm i} \nu_m)
\nonumber \\
&&  \hspace{20mm}
\times \Phi^{LL'} ({\bf k}+{\bf q}'; {\rm i} \omega_n+ {\rm i} \nu_m)
\bigg\},
\label{eq11}
\end{eqnarray}
%
%
in obvious notation ($E_{LL'} ({\bf k},{\bf k}_+) = 
E_{L} ({\bf k}) - E_{L'} ({\bf k}_+)$  is the electron-hole-pair energy).
	In the intraband channel, these equations transform into two 
coupled equations, one is the charge continuity equation and the other is 
the transport equation.
	These two equations have to be solved self-consistently.
	For this purpose, we retain in Eqs.~(\ref{eq11})
only the most singular scattering processes  by putting  
${\rm i} \omega_{n+}  \rightarrow E_{L'} ({\bf k}_+)/\hbar$ and 
${\rm i} \omega_{n}  \rightarrow E_{L} ({\bf k})/\hbar$, in two single-particle
self-energies, $\Sigma_{L} ({\bf k}, {\rm i} \omega_n )$ 
and $\Sigma_{L'} ({\bf k}_+, {\rm i} \omega_{n+})$, and in the 
related vertex corrections on the right-hand side of the equations.
	In this way, the expression in the brackets on the left-hand side 
does not depend on ${\rm i} \omega_n$, and the equations can be rewritten 
in terms of the intraband ($L=L'$) and interband ($L \neq L'$)
induced densities $\delta n^{LL'} ({\bf k},\omega)$
defined by 
\begin{eqnarray} 
\delta n^{LL'} ({\bf k},\omega)
&=& \sum_{{\bf k}'}  \frac{1}{\hbar}
{\cal D}^{LL'} ({\bf k}, {\bf k}_+,{\bf k}'_+,{\bf k}',\omega)
\nonumber \\
&& \times
eq^{L'L} ({\bf k}_+,{\bf k}) V^{\rm ext} ({\bf q},\omega).
\label{eq12}
\end{eqnarray}
%
%
	Here
\begin{eqnarray} 
eq^{L'L} ({\bf k}_+,{\bf k}) V^{\rm ext} ({\bf q},\omega) \equiv 
\frac{\hbar  J_{\alpha}^{L'L} ({\bf k})}{
E_{L'L}({\bf k}_+,{\bf k})} \, {\rm i} E_{\alpha} (\omega),
\label{eq13}
\end{eqnarray}
%
%
with $E_{\alpha} (\omega)$ and 
$V^{\rm ext} ({\bf q},\omega) = ({\rm i}/q_\alpha) E_{\alpha} (\omega)$
being, respectively, the external electric field and the 
related scalar potential (notice that for $L=L'$ Eq.~(\ref{eq13}) reduces 
to the expression 
$eV^{\rm ext} ({\bf q},\omega) = e({\rm i}/q_\alpha) E_{\alpha} (\omega)$
that is independent of {\bf k}).  
	Furthermore, $q^{L'L} ({\bf k}_+,{\bf k})$ is the charge vertex, given
by Eqs.~(\ref{eqA8}) and (\ref{eq29}),
$J^{L'L}_{\alpha} ({\bf k})$ is the current vertex 
and ${\bf q} =  q_{\alpha} \hat{e}_{\alpha}$.	

In Eq.~(\ref{eq12}),
${\cal D}^{LL'} ({\bf k}, {\bf k}_+,{\bf k}'_+,{\bf k}',\omega)$ 
is obtained by the analytical continuation of 
${\cal D}^{LL'} ({\bf k}, {\bf k}_+,{\bf k}'_+,{\bf k}',{\rm i} \nu_n)$,
${\rm i} \nu_n \rightarrow \omega + {\rm i} \eta$.
	The summation  of Eqs.~(\ref{eq11}) over ${\rm i} \omega_n$ 
makes the single-particle self-energy contributions and 
the related vertex corrections to appear  on the same  footing,
resulting in  the electron-hole  self-energy of the  form
\begin{eqnarray} 
&& \! \!\! \!\! \!\! \!
\hbar {\it \Pi}^{LL'}_{\alpha} ({\bf k}, \omega) 
\label{eq14}\\
&&\approx 
-\sum_{\lambda {\bf q}'} \sum_{s = +1, -1} 
\frac{\big| G_\lambda({\bf q}')\big|^2}{N}
\bigg\{1 - \frac{\tilde J_{\alpha}^{LL'}({\bf k}+{\bf q}')}{
\tilde J_{\alpha}^{LL'}({\bf k})} \bigg\}
\nonumber \\ 
&& \hspace{5mm} \times
\bigg[ \frac{f^{\rm b}_\lambda({\bf q}') + f(sE_{L'}({\bf k}+ {\bf q}'))}{
\hbar \omega +{\rm i} \eta + E_{LL'} ({\bf k}, {\bf k}+{\bf q}') + 
s\hbar \omega_{\lambda {\bf q}'}}
\nonumber \\ 
&&  \hspace{10mm}
+\frac{f^{\rm b}_\lambda({\bf q}') + f(sE_{L}({\bf k}+ {\bf q}'))}{
\hbar \omega +{\rm i} \eta + E_{LL'} ({\bf k}+{\bf q}', {\bf k}) - 
s\hbar \omega_{\lambda {\bf q}'}} \bigg]. 
\nonumber
\end{eqnarray}
%
%
	The corresponding generalization of 
${\it \Pi}^{LL'}_{\alpha} ({\bf k},  \omega)$ to the case with the interband 
electron-phonon coupling is straightforward; however, it is beyond the 
scope of the present work.
	$f^{\rm b}_\lambda({\bf q}')$ and $f(E_L({\bf k}))\equiv f_L({\bf k})$ 
in Eq.~(\ref{eq14}) 
are, respectively, the Bose--Einstein and Fermi--Dirac distribution functions.
	Finally, notice that the electron scattering on the static disorder, 
described by the  potential $V_1({\bf q}')$, 
is included here  through the replacement
$\omega_{\lambda {\bf q}'} \rightarrow 0$ and
$(1+2 f^{\rm b}_\lambda({\bf q}')) |G_\lambda({\bf q}')|^2/N 
\rightarrow |V_1({\bf q}')|^2$ in the $\lambda=0$ boson branch 
\cite{KupcicPRB}.

\subsection{Transport equations and optical conductivity}

The resulting self-consistent equation for the densities \\
$\delta n^{LL'} ({\bf k},\omega)$
represents the generalization of the transport equation
\cite{Pines,KupcicPRB},
\begin{eqnarray}
&& \! \! \! \!\! \! \! 
\big[ \hbar \omega + E_{LL'}({\bf k},{\bf k}_+) \big]
\delta n^{LL'} ({\bf k},\omega) 
+ \hbar {\it \Pi}^{LL'}_{\alpha} ({\bf k}, \omega)
\delta \tilde{n}^{LL'} ({\bf k},\omega)
\nonumber \\
&& \hspace{10mm} 
= \frac{f_L({\bf k}) - f_{L'}({\bf k}_+) }{
E_{L'L}({\bf k}_+,{\bf k})} \,
{\rm i}\hbar  J_{\alpha}^{L'L} ({\bf k}) E_{\alpha} (\omega). 
\label{eq15}
\end{eqnarray}
%
%
	In this equation  $\delta \tilde{n}^{LL'} ({\bf k},\omega)$ 
is the contribution  to $\delta n^{LL'} ({\bf k},\omega)$ proportional 
to $\tilde J_{\alpha}^{L'L} ({\bf k})$
(the effective current vertices in Eqs.~(\ref{eq14})  and (\ref{eq15}) are 
$\tilde J_{\alpha}^{LL} ({\bf k}) = J_{\alpha}^{LL} ({\bf k})$
in the intraband channel and 
$\tilde J_{\alpha}^{L\underline{L}} ({\bf k}) 
= \hbar \omega J_{\alpha}^{L\underline{L}} ({\bf k})/
E_{L\underline{L}} ({\bf k},{\bf k})$ in the interband channel).
	Equation (\ref{eq15}) is solved consistently with the 
charge continuity  equation and combined with the definition
of the optical conductivity tensor.
	The result is the optical conductivity 
of the effective single-particle multiband models
\cite{KupcicPB1,KupcicPRB,KupcicPB2,Lynge}
\begin{eqnarray}
&& \! \! \! \! \! 
\sigma_{\alpha \alpha}^{\rm sp} (\omega) = 
\frac{\rm i}{\omega} \frac{1}{V} \sum_{LL' {\bf k} \sigma}
\bigg( \frac{\hbar \omega }{E_{L'L}({\bf k}_+,{\bf k}) } \bigg)^{n_{LL'}}
\big| J_{\alpha}^{LL'} ({\bf k}) \big|^2
\label{eq16} \\ \nonumber 
&& \hspace{10mm} \times
\frac{f_L({\bf k}) - f_{L'}({\bf k}_+)}{
\hbar \omega + \hbar {\it \Pi}^{LL'}_{\alpha} ({\bf k}, \omega)
+ E_{LL'}({\bf k},{\bf k}) 
\displaystyle -\frac{E^2_{L'L'}({\bf k},{\bf k}_+)}{\hbar \omega}},
\end{eqnarray}
%
%
with $n_{LL} = 1$ in the intraband channel and $n_{L\underline{L}} = 2$ 
in the interband channel. In the dynamical limit, the last term in the 
denominator can be safely neglected.
	At this point it should be recalled that the current vertices 
$J^{LL'}_{\alpha} ({\bf k})$ can be shown in simple physical terms as:
$J^{LL}_{\alpha} ({\bf k}) = e v^L_{\alpha} ({\bf k})$ 
in the intraband channel and
$J_{\alpha}^{L\underline{L}} ({\bf k}) = ({\rm i}/ \hbar) 
P_{\alpha}^{L\underline{L}} ({\bf k}) E_{L\underline{L}} 
({\bf k}, {\bf k})$ in the interband channel;
$v^L_{\alpha} ({\bf k})$ is the electron group velocity and
$P_{\alpha}^{L\underline{L}} ({\bf k})$ is the related interband dipole vertex.
	It should also be noticed that the coherence factor 
in the textbook expression for the interband contributions \cite{LRA,Gruner} 
is slightly different from that in Eq.~(\ref{eq16}), as well as that the 
phenomenological extension of the textbook expression is found to lead to
the non-physical in-gap spectra (see Fig.~1 in Ref.~\cite{Kim}).

\subsection{$\Delta =0$ case}
%
%
Before turning to the detailed numerical analysis of the expression (\ref{eq16}), 
let us write Eqs.~(\ref{eq15}) and (\ref{eq16}) explicitly for the (single-band) case in which 
$\Delta =0$ and the relaxation rate $\Gamma$ is presumably small but finite
(${\rm Im} \{ {\it \Pi}^{0}_{\alpha} ({\bf k}, \omega)\} \approx \Gamma$).
	These expressions will make the comparison of the present approach
with the common CDW approaches in Sec.~5 straightforward.

For $\Delta =0$ and ${\bf q} = q_x \hat{e}_x$, the electron-hole-pair energy is
$\varepsilon({\bf k}_+) - \varepsilon({\bf k}) \approx \hbar \omega_0 ({\bf q})
= \hbar q_x v^0_{x} ({\bf k})$, with 
$\omega_0^2 ({\bf q})$ being independent  of {\bf k} for 
${\bf k}\approx {\bf k}_{\rm F}$. 
	Eq.~(\ref{eq15}) can be written now in the textbook form \cite{Pines}
\begin{eqnarray}
&& \! \! \! \!\! \! \!
\omega \delta n_0 ({\bf k},\omega)
-  \omega_0 ({\bf q}) \delta n_1 ({\bf k},\omega) = 0,
\nonumber \\
&& \! \! \! \!\! \! \!
\big( \omega + {\rm i}\Gamma \big)\delta n_1 ({\bf k},\omega)
-  \omega_0 ({\bf q}) \delta n_0 ({\bf k},\omega) 
 \nonumber \\ 
 && \hspace{25mm}
= (-)\frac{\partial f({\bf k}) }{\partial \varepsilon ({\bf k})}\,
{\rm i}  ev_{x}^{0} ({\bf k}) E_{x} (\omega). 
\label{eq17}
\end{eqnarray}
%
%
	Introducing the boson field $\psi ({\bf q},\omega)$, which time
derivative is proportional to the current induced by the external electric 
field,
\begin{eqnarray}
j_x^{\rm ind} (\omega) =
\frac{1}{V} \sum_{{\bf k} \sigma} ev^0_{x} ({\bf k}) 
\delta n_1 ({\bf k},\omega)  \equiv \frac{e}{\pi}\dot{\psi} ({\bf q},\omega),   
\label{eq18}
\end{eqnarray}
%
%
one obtains the equation of motion
\begin{eqnarray}
m\big[ \omega \big( \omega + {\rm i}\Gamma \big)
-  \omega^2_0 ({\bf q}) \big]\psi ({\bf q},\omega) 
&=&  e \pi n^{\rm eff,0}_{xx} E_{x} (\omega). 
\label{eq19}
\end{eqnarray}
%
%
	The resulting optical conductivity is given by 
the interband term in Eq.~(\ref{eq16}), i.e., by the expressions
\begin{eqnarray}
\sigma_{xx} (\omega) &=& 
\frac{e^2 n^{\rm eff,0}_{xx}}{m}
\frac{{\rm i}  \omega}{\omega\big(  \omega +  {\rm i}\Gamma \big)  
-\omega^2_0({\bf q})} 
\nonumber \\
&\equiv& \frac{{\rm i}  \omega}{\hbar}
\bigg( \frac{e}{\pi}  \bigg)^2 V {\cal D}^0_\psi ({\bf q},\omega).
 \label{eq20} 
\end{eqnarray}
%
%
	Here ${\cal D}^0_\psi ({\bf q},\omega)$ is the bare propagator of the 
field $\psi ({\bf q},\omega)$, $n^{\rm eff,0}_{xx}$ is the effective
number of conduction electrons,
\begin{equation}
n^{\rm eff,0}_{xx} =
-\frac{m}{V} \sum_{{\bf k} \sigma} 
\big(v^0_{x} ({\bf k}) \big)^2
\frac{\partial f({\bf k}) }{\partial \varepsilon({\bf k})}
= \frac{2}{\pi} \frac{m_{aa} v_x^0({k}_{\rm F})}{\hbar},
\label{eq21}
\end{equation}
%
%
$v^0_{x} ({\bf k}) = (1/\hbar) \partial \varepsilon({\bf k})/\partial k_x$ 
is the bare electron group velocity and 
$m_{aa} = 2 \hbar^2/(t_a a^2)$ is the mass parameter.


\section{Discussion}
%
\subsection{Underdamped regime}
%

For comparison with the textbook expressions for the 
single-particle contributions to the optical conductivity, 
it is convenient to turn back to the simplest case discussed in Sec.~3.1. 
	The electron-hole self-energy is approximated 
by its imaginary part and the imaginary part is assumed to be 
small when compared to the interband  threshold energy; i.e.,
${\it \Pi}^{LL'}_{\alpha} ({\bf k},\omega) \approx 
{\rm i}\Gamma^{LL'}_{\alpha} ({\bf k}, \omega)$
and
$\Gamma^{LL'}_{\alpha}({\bf k},\omega) \approx \Gamma^{LL'}_{\alpha}$
($\Gamma^{LL}_{\alpha} \equiv \Gamma_{\alpha}^{\rm intra}$ and
$\Gamma^{L\underline{L}}_{\alpha} \equiv \Gamma_{\alpha}^{\rm inter}$,
hereafter).
	In the underdamped regime ($\Gamma_{\alpha}^{\rm inter} \ll 2\Delta$), 
this approximation does not affect
the spectrum in a critical manner, neither
in the metallic regime (with $2k_{\rm F}$  not too close to $Q_x$),
nor  in the CDW state at temperatures well below the critical temperature.
	In the  metallic case, the
effective numbers of conduction electrons in the resistivity,
or in the Drude part of the optical conductivity, consist of two parts,
the residual one, which remains finite at $T \rightarrow 0$, and
the thermally activated part.
	In the $t_b \rightarrow 0$ Peierls case
there are only thermally activated electrons.
	Finally, in the Peierls case with $t_b / \Delta$ not too small there 
is a portion of the Fermi surface on which the CDW gap is not developed,
leading again to a finite residual number of conduction electrons.

\begin{figure}[tb]
\begin{center}
\resizebox{0.3\textwidth}{!}{%
  \includegraphics{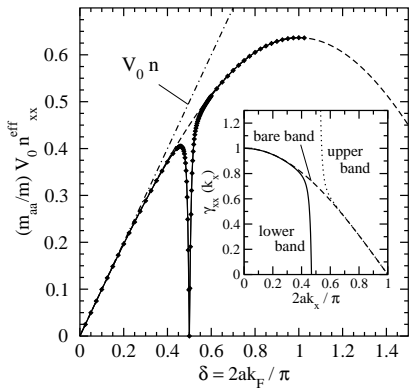}
  }    
    \caption{
    Main figure: 
    The normalized effective number of conduction electrons 
    $(m_{aa}/m) V_0 n^{\rm eff}_{xx}$   as a function of the 
    band filling for $2t_a = 0.5$ eV, $2t_b =0$, 
    $T \rightarrow 0$ and $\Delta_0 = 0$ 
    (dashed line) and 10 meV (solid line).
    The diamonds represent the dimensionless electron group velocity
    $[2/(t_a \pi^2)]v^0_{x} (k_{\rm F})$.
    Inset:
        The  inverse effective mass tensors
    $\gamma_{xx}^{CC} ({\bf k})$ and 
    $\gamma_{xx}^{\underline{C}\underline{C}} ({\bf k})$ 
    for $Q_x = 0.5 \pi/a$ and $\Delta_0 =  10$ meV
    (solid and dotted lines, respectively).
    The bare-band vertex ($\Delta=0$, dashed line) is given for
    comparison.
    }
\end{center}    
\end{figure}

To compare briefly the correct estimations  of these numbers with
the oversimplified ``semiconducting'' expressions encountered 
in the literature \cite{Banik}, we apply Eq.~(\ref{eq16}) now to  
the simplest metallic situation corresponding to
${\bf Q} = (0.5\pi/a,0)$ and $t_b \rightarrow 0$. 
 	In this respect, it should be noticed  that the present analysis  
is not focussed on the single-particle properties of the usual strictly 1D model,
the physics of which is dominated by the singular nature of the 
forward scattering processes in the single-electron self-energy.
	Instead, we consider the electron-hole excitations in the typical 
Q1D model in which the Q1D character is incorporated 
in the aforementioned form of the electron-hole self-energy,
where the forward scattering contributions are removed by
the Ward identity cancellation of the ${\bf q} \approx 0$ 
self-energy and vertex corrections in Eq.~(\ref{eq14}).
 	The $t_b \rightarrow 0$ limit serves here only to replace 
most of the integrations over the 2D Brillouin zone by the 1D integrations.
	The results obtained in this way can be compared term by term
to the well-known analytical expressions.

For example, the  intraband conductivity  
along the highly conducting direction
is proportional to the effective number of conduction electrons
$n^{\rm eff}_{xx}$, as seen from 
\begin{eqnarray}
\sigma_{\alpha \alpha}^{\rm intra}  (\omega) \approx 
\frac{{\rm i}e^2 n^{\rm eff}_{\alpha \alpha}}{ 
m(\omega + {\rm i} \Gamma^{\rm intra}_\alpha)},
\label{eq22}
\end{eqnarray}
%
%
for $\alpha = x$.
	$n^{\rm eff}_{\alpha \alpha}$ includes  the contributions of both bands,
and it can be shown in two alternative ways, as a surface integral in the 
{\bf k} space, or the related volume integral,
\begin{eqnarray}
n^{\rm eff}_{\alpha \alpha} \hspace{2mm} =  \hspace{2mm}
\sum_L n^{\rm eff}_{\alpha \alpha} (L) 
&=&
-\frac{m}{V} \sum_{L{\bf k} \sigma} 
\big(v^L_{\alpha} ({\bf k}) \big)^2
\frac{\partial f_L({\bf k}) }{\partial E_L({\bf k})}
\nonumber \\
&=& \frac{1}{V} \sum_{L{\bf k} \sigma} 
\gamma^{LL}_{\alpha \alpha} ({\bf k}) f_L({\bf k}).
\label{eq23}
\end{eqnarray}
%
%
	In the first representation, it is given in terms of 
the electron group velocity,
while in the second representation it depends on 
the dimensionless inverse effective-mass tensor
$\gamma^{LL}_{\alpha \alpha} ({\bf k}) = 
(m/\hbar) \partial v^L_{\alpha} ({\bf k})/ \partial k_\alpha$.
	The dependence of $n^{\rm eff}_{xx}$  on the 
electron doping $\delta = 2a k_{\rm F} /\pi$ is shown in Fig.~4
for two choices of the distortion potential,  
$\Delta_0 =0$ and 10 meV, 
and $T \rightarrow 0$.
	The electron doping $\delta = nV_0$ 
($n$ is the concentration of electrons and $V_0$ is the primitive cell volume
of the $\Delta_0 =0$ lattice)	
and the  dimensionless electron group velocity are also given for comparison.
	The latter represents the analytical solution of the 
$t_b \rightarrow 0$ model.

\begin{figure}[tb]
\begin{center} 
\resizebox{0.3\textwidth}{!}{%
  \includegraphics{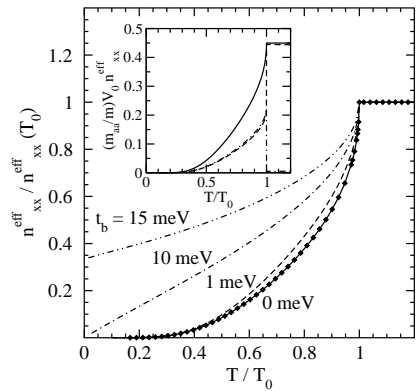}}    
        \caption{Main figure: The  temperature dependence of the 
    normalized effective number of conduction electrons 
    in the (imperfectly nested $Q_y=0$) CDW state for 
    $2t_a = 0.5$ eV, $2t_b =0$ (solid line), 1 meV (dashed line),
    10 meV (dot-dashed line) and 15 meV (dot-dot-dashed line), 
    $2k_{{\rm F}} = Q_x  = 0.5 \pi/a$, $T_{0} =66.7$ K and
    $\Delta_0 = 10$ meV.
    The diamonds represent the function $\exp \{ -\Delta(T)/(k_{\rm B} T) \}$.
    }
\end{center} 
\end{figure}

The metallic regime of the primary experimental interest corresponds
to the case of a low electron/hole doping with respect to the
quarter-filled band.
	In this case, the residual part in $n^{\rm eff}_{xx}$ 
(shown in Fig.~4) is  small when compared to its 
$\Delta =0$ value (dashed line in the figure).
	Even for relatively low  temperatures, the thermally activated
part in $n^{\rm eff}_{xx}$ becomes important, leading to pronounced
thermal effects in $n^{\rm eff}_{xx}$.
	The thermal effects in the Drude part of the optical conductivity,
or in the resistivity, are further illustrated in Fig.~5 for the 
imperfectly nested CDW case  
(corresponding to the electron doping $\delta = 0.5$ in Fig.~4). 
	The temperature dependence of 
the activation energy  $\Delta $ is  parametrized 
as $\Delta (T) = \Delta_0 (1- T/T_0)^\beta$, with $\beta = 1/2$.
	$n^{\rm eff}_{xx}$ is shown for four different values of $t_b$ 
and compared to the empirical expression 
$n^{\rm eff}_{xx}(T) = \exp \{ -\Delta(T)/(k_{\rm B} T) \}
n^{\rm eff}_{xx}(T_0)$
(the residual part in $n^{\rm eff}_{xx}$ corresponds to the 
zero-temperature interception value).
	The dashed and dot-dashed lines in the inset of figure
represent the partial contributions of the lower and upper band for 
$t_b =0$,
$n^{\rm eff}_{xx}(C)$ and $n^{\rm eff}_{xx}(\underline{C})$ in Eq.~(\ref{eq23}),
respectively.
	The latter two have to be contrasted to the numbers 
\begin{eqnarray}
n_{\rm h} = \frac{1}{V}\sum_{{\bf k} \sigma} [1-f_C({\bf k})],
&&
n_{\rm e} = \frac{1}{V}\sum_{{\bf k} \sigma} f_{\underline{C}}({\bf k})
\label{eq24}
\end{eqnarray}
%
%
invoked from the theory of the  ordinary semiconductors.
	The numbers (\ref{eq24}) are often taken as a  good approximation for 
$n^{\rm eff}_{xx}(C)$ and $n^{\rm eff}_{xx}(\underline{C})$.
	However, the disagreement between the
expressions (\ref{eq24}) and experimental observation is typically 
of two orders of magnitude \cite{Forro},
and, consequently,  the analyses  based on the numbers (\ref{eq24})
have to be taken with  reservation.
	For the parameters used in Fig.~5, we obtain a factor of 25.
	This disagreement is easily understood on noting that the 
semiconducting approach, Eq.~(\ref{eq24}), completely neglects 
the coherence effects in the inverse effective-mass tensor.
 	In other words, by replacing the vertices $\gamma^{LL}_{xx} ({\bf k})$ 
by unity, when the numbers $n^{\rm eff}_{xx}(C)$ and 
$n^{\rm eff}_{xx}(\underline{C})$  reduce to $n_{\rm h}$ and $n_{\rm e}$ indeed,
this approach does not take carefully into account the fact that 
the thermally activated electrons 
are related only to the states in the thermal window around the Fermi level. 
	Notice in this respect the singular behavior of
the vertices $\gamma^{LL}_{xx} ({\bf k})$ at $2k_x = 0.5 \pi/a$, shown in the 
inset of Fig.~4, which dominates the behavior of 
$n^{\rm eff}_{xx}$ for $2k_{\rm F} \approx 0.5 \pi/a$.
	On the contrary, the common approaches based on 
the surface-integral  representation of $n^{\rm eff}_{\alpha \alpha}$
usually give rise to the  correct results.

\begin{figure}[tb]
\begin{center}
\resizebox{0.3\textwidth}{!}{%
  \includegraphics{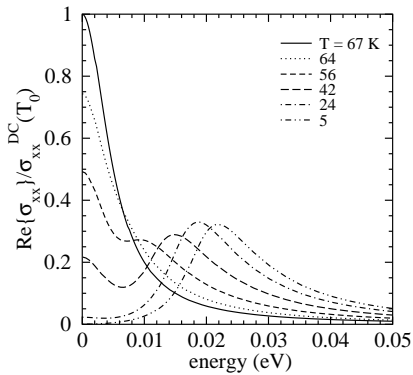}}           
    \caption{
    The development of the normalized total optical 
    conductivity with temperature in the underdamped regime:
    $2t_a = 0.5$ eV, $2t_b =1$ meV, 
    $T_{0} =66.7$ K, $\Delta_0 =10$ meV
    and  $\hbar \Gamma^{\rm intra}_a = \hbar \Gamma^{\rm intra}_a = 5$ meV.
    }
\end{center}
\end{figure}

Another important observation is that the optical conductivity model 
(\ref{eq16}) has a quite general form and is thus
applicable to variety of low-dimensional systems.
	In this respect one should notice that 
the values of the effective numbers $n^{\rm eff}_{\alpha \alpha}$ are generally
a direct consequence of the coherent scattering of conduction 
electrons on the crystal potential or
on the periodically distorted lattice, irrespective of the band filling.
	These numbers represent the numbers of effectively free charge
carriers, with  the thermally activated part being  usually very sensitive to
temperature, providing  the estimation of the activation energy $\Delta$, 
or the charge-transfer gap in the general case. 
	The corresponding numbers of bound charges, on the other hand,  
are  the numbers of excited electrons across the CDW gap \cite{Gruner}
or across the charge-transfer gap. 
	The latter numbers depend on temperature as well, as seen from the 
interband part of Eq.~(\ref{eq16}), which can be rewritten in the form
\begin{eqnarray}
\sigma_{\alpha \alpha}^{\rm inter} (\omega) 
&\approx& 
{\rm i}\omega \frac{1}{V} \sum_{L\neq L' {\bf k} \sigma}
\big| P_{\alpha}^{LL'} ({\bf k}) \big|^2
\nonumber \\ 
&& \times 
\frac{  f_L({\bf k}) -f_{L'}({\bf k})}{
\hbar \omega + {\rm i}\hbar \Gamma_\alpha^{\rm inter} 
+ E_{LL'}({\bf k},{\bf k})}.
\label{eq25}
\end{eqnarray}
%
%

The temperature dependence of the total optical conductivity 
$\sigma_{xx}^{\rm sp} (\omega) = \sigma_{xx}^{\rm intra} (\omega)
+ \sigma_{xx}^{\rm inter} (\omega)$
is calculated for  several  underdamped  regimes.
	The damping energies $\hbar \Gamma^{\rm intra}_a$ and 
$\hbar \Gamma^{\rm inter}_a $ are assumed to be temperature independent,
for simplicity, and small in comparison with the scale $2\Delta$.
	The results for the imperfectly nested case are shown in Fig.~6 and 
are essentially the same as the results of the 
analogous nearly half-filled dimerized band \cite{KupcicPB1,Lynge}.
	Although it is physically clear that at temperatures $T$ well below 
$T_{\rm c}$ $\Gamma^{\rm inter}_a $ is much larger than $\Gamma^{\rm intra}_a$, 
the spectra shown in Fig.~6 are calculated for 
$\Gamma^{\rm intra}_a = \Gamma^{\rm inter}_a $, for clarity. 
	The shift of the MIR maximum in the spectra 
($\hbar \omega_{\rm MIR}$) with temperature
follows  roughly the energy scale $2\Delta$, giving an alternative
method of estimating  $\Delta$ \cite{LRA}.
	The spectra are also characterized by the in-gap optical excitations, 
mainly in the energy range from 
$\hbar \omega_{\rm min} \approx 2\Delta - \hbar \Gamma^{\rm inter}_a$
to $ 2\Delta$.
	The  cut-off energy  $\hbar \omega_{\rm min}$ is associated with 
the zero conductivity interception of the line obtained by the linear
extrapolation of the interband conductivity.
The square-root singularity at $\hbar \omega = 2 \Delta$ is absent here, and,
according to Ref.~\cite{McKenzie}, the same shape of 
$\sigma_{xx}^{\rm inter} (\omega)$ is expected down to zero temperature.

\begin{figure}[tb]
\begin{center} 
\resizebox{0.3\textwidth}{!}{%
  \includegraphics{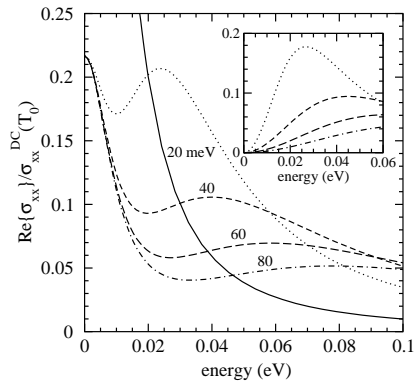}}        
    \caption{
    Main figure: The optical conductivity as a function of 
    damping energy $\hbar \Gamma^{\rm inter}_a$ at $T = 42$ K.
    Dotted, dashed, long-dashed and dot-dashed lines correspond to
    $\hbar \Gamma^{\rm inter}_a = 20$, 40, 60 and 80 meV,
    respectively.
    The other parameters are the same as in Fig.~6.
    The solid line in the total optical conductivity at $T = 67$ K.
    Inset: The related interband contributions 
    (the parameters are the same as in the main figure).
    }
\end{center}
\end{figure}  

\subsection{Overdamped regime}
%

A completely different situation appears in the overdamped regime
where the damping energy $\hbar \Gamma^{\rm inter}_a $ becomes comparable to
$2\Delta$ and where the real part of the electron-hole self-energy starts 
playing an important role.
	Typical physical situation corresponding to this regime is the 
CDW instability of the magnetic metals, where the wide band electrons
exhibit the CDW instability, and are usually scattered by the strongly
correlated electrons in one or more narrow bands.
	As the present discussion of the general expression (\ref{eq16})  
neglects the real part of the self-energy,
it turns out to be  inappropriate  for the detailed quantitative analysis
of the overdamped regime, but it satisfactorily explains the interplay 
between two mutually competing energy scales, $\hbar \Gamma^{\rm inter}_a $ 
and $2\Delta$.  
	Fig.~7 shows typical results.
	The position of the MIR peak is at 
$\hbar \omega_{\rm MIR} \approx 
\sqrt{4 \Delta^2 + (\hbar \Gamma^{\rm inter}_a)^2}$.
	In this case, $\hbar \omega_{\rm min}$ is a rather complicated function of the 
model parameters, but it is close to the value obtained in the 
narrow band limit of $\sigma_{xx}^{\rm inter} (\omega)$, where
all interband excitations are assumed to appear close to the threshold energy
$E_{\underline{C}C} ({\bf k},{\bf k}) \approx 2 \Delta$.
	In this case, one obtains
\begin{eqnarray}
{\rm Re} \{ \sigma_{xx}^{\rm inter} (\omega) \} &\propto &
\frac{\hbar \omega  \hbar\Gamma^{\rm inter}_a}{
(\hbar\omega - 2 \Delta)^2 + (\hbar\Gamma^{\rm inter}_a)^2}  
\nonumber \\ 
&&\times
\frac{8 \hbar \omega \Delta}{
(\hbar\omega + 2 \Delta)^2 + (\hbar \Gamma^{\rm inter}_a)^2}. 
\label{eq26}
\end{eqnarray}
%
%
	As a result, in the overdamped regime
$\hbar \omega_{\rm min}$ increases  with increasing 
$\hbar \Gamma^{\rm inter}_a $, for example, from the value 
$\hbar \omega_{\rm min} \approx \Delta$ at 
$\hbar \Gamma^{\rm inter}_a = 5 \Delta$ to
$\hbar \omega_{\rm min} \approx 2\Delta$ at 
$\hbar \Gamma^{\rm inter}_a = 10 \Delta$.
	Knowing the experimental values of $\hbar \omega_{\rm MIR}$
and $\hbar \omega_{\rm min}$, we can again estimate $\Delta$ 
and $\hbar \Gamma^{\rm inter}_a$ and compare $\Delta$  
to the activation energy of transport coefficients.

As mentioned in the introduction, a clear evidence of the overdamped regime 
is recently found  in the  vanadium chain compound BaVS$_3$ \cite{Kezsmarki}. 
	 BaVS$_3$ is the multiband system with the (commensurate) 
CDW instability in which one weakly correlated wide band and
two strongly correlated narrow bands coexist at the Fermi level.
	Thanks to the extensive experimental activity 
in this system in recent years \cite{Forro00,Fagot,Kezsmarki,Mitrovic}, 
BaVS$_3$ seems to be a good candidate 
for detailed investigation of the overdamped regime.

\section{Comparison with the common CDW theories}
%
%

In the low-energy examinations of the  CDW systems it is common 
to use different strictly 1D models with the non-retarded short-range
interactions in which not only the forward-scattering but also the 
backward-scattering processes in the electron-hole self energy are neglected
\cite{Solyom,Giamarchi}.
	These approaches are usually  based on the bosonization procedure.
	For example, for the simplest case with only the forward-scattering
interaction $g_4$ taken into account, the optical conductivity 
is given by the RPA diagrams in Fig.~8(a).
	The first diagram is nothing but the ideal intraband 
conductivity of Eq.~(\ref{eq20}).
	The interaction $g_4$ leads to the renormalization of the electron group
velocity in Eqs.~(\ref{eq20}) and (\ref{eq21}) in the way that
$$
v_x^0({\bf k}) \rightarrow u_\rho =  \sqrt{\frac{m}{m^*}}v_x^0(k_{\rm F}),
$$ 
while the electron dispersions in 
$\partial f({\bf k})/\partial \varepsilon ({\bf k})$,
together with the related density of states, remain unaffected by $g_4$.
	The boson field $\psi({\bf q}, \omega)$ is found to satisfy 
the equation of motion (\ref{eq19}), while the optical conductivity 
is found to be given by Eqs.~(\ref{eq20}), (\ref{eq21}), 
with  $v_x^0({\bf k}) \rightarrow u_\rho$ \cite{Fukuyama,Giamarchi}.

The present Q1D model  differs from these strictly 1D models 
in three important points.
	First, the electron-electron interactions  in question are
the retarded, phonon-mediated interactions \cite{LRA,Schulz,Rice},
second, the collective infrared-active mode is the phase phonon mode,
which participates in the total conductivity spectral weight with 
usually less than 0.5 \% (e.g., $m^*$ is of the order of $400m$ 
in K$_{0.3}$MoO$_3$ \cite{Degiorgi}), 
and, finally, the external electric fields are assumed to be
well below the critical field required to give the non-linear conductivity
\cite{Artemenko,Maki}.

 \begin{figure}[tb]
\begin{center}
\resizebox{0.45\textwidth}{!}{%
\includegraphics{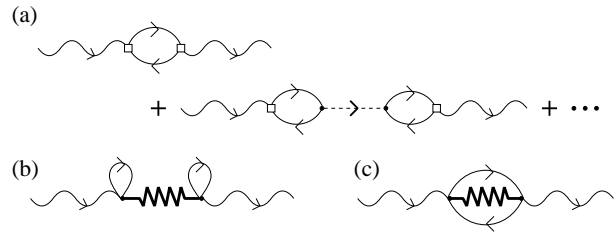}}
 \caption{
(a) The collective contribution to the optical conductivity in 
the common 1D models with the forward-scattering short-range interactions. 
(b) The collective contribution in the present model.
(c) The Hopfield-like processes neglected in the present approach.
}
\end{center}
\end{figure}

In this section, we want to determine the structure of the 
photon-phase phonon coupling, and to compare the 
obtained results with both the  textbook results obtained within 
the same Q1D model \cite{LRA,Schulz,Rice,Gruner}
and  the results of strictly 1D models \cite{Fukuyama,Giamarchi}.
	We will derive here, for the first time, a clear selection rule 
for the infrared-active (and Raman-active) phonon modes.
	On the qualitative level, the arguments are the following.	
For the  quarter-filled case shown in Fig.~1, 
the electric fields couple to the dipole moment of the 
phase phonon mode (the dipole moment is carried by the bound (condensed) 
electrons).
	For $\Delta$ comparable to the band width, this mode is the ordinary
infrared-active  phonon mode, characterized by the effective ion charge $q =
q_c/\sqrt{2}$, the reduced ion mass $M$ and the concentration of ions 
$1/V_0$.
	On the other hand, for  $\Delta$  small in comparison with the band 
width, the electrons in question are only weakly bound and the magnitude of
the dipole  is now proportional to $\tilde{e}$,  which is $q$ multiplied by
the ratio between the amplitudes of the electron and ion oscillations. 
	In the limit $\Delta \rightarrow 0$, the dipole moment vanishes.


 \begin{figure}[tb]
\begin{center}
\resizebox{0.45\textwidth}{!}{%
  \includegraphics{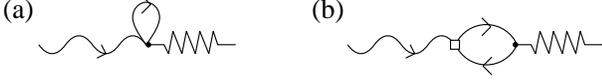}}
 \caption{
(a) The direct photon-phonon coupling and (b) the indirect, electron-mediated 
photon-phonon coupling in pure CDW systems. 
    }
\end{center}
\end{figure}

To discuss this question quantitatively, 
we neglect again the Umklapp processes in the electron dispersions 
and focus the attention on the $\alpha = 0$ two-band model
of Eq.~(\ref{eq4}).
	In this case, all relevant vertex functions are simple functions 
of the auxiliary phase $\varphi ({\bf k})$ defined by
\begin{eqnarray}
\tan \varphi ({\bf k}) = \frac{2\Delta}{\varepsilon_{
\underline{c}c} ({\bf k})}. 
\label{eq27}
\end{eqnarray}
%
%
	In the long wavelength limit, the electron-phonon vertices  and the 
charge vertices of Eqs.~(\ref{eqA5}) and (\ref{eqA8})  become	
\begin{eqnarray}
g^{CC}_A ({\bf k}_+, {\bf k}) &=& 
-g^{\underline{C}\underline{C}}_A ({\bf k}_+, {\bf k})
\nonumber \\
&=&  -\sin \frac{\varphi ({\bf k}_+)+\varphi ({\bf k})}{2} \approx 
- \sin \varphi ({\bf k}), 
\nonumber \\
g^{\underline{C}C}_A ({\bf k}_+, {\bf k}) &\approx& 
{\rm e}^{{\rm i} \phi ({\bf k})} \cos \varphi ({\bf k}),
\nonumber \\
g^{CC}_\varphi ({\bf k}_+, {\bf k}) &=&  
g^{\underline{C}\underline{C}}_\varphi ({\bf k}_+, {\bf k})
\nonumber \\
&=& {\rm i} \sin \frac{\varphi ({\bf k})-\varphi ({\bf k}_+)}{2}
\approx {\cal O}(q_\alpha), 
\nonumber \\
g^{\underline{C}C}_\varphi ({\bf k}_+, {\bf k}) &\approx&  
{\rm i} {\rm e}^{{\rm i} \phi ({\bf k})}, 
\label{eq28} \\
q^{CC} ({\bf k}_+, {\bf k}) &\approx&
q^{\underline{C}\underline{C}} ({\bf k}_+, {\bf k})
\approx 1, 
\nonumber \\
q^{\underline{C}C} ({\bf k}_+,{\bf k}) &\approx&  
\frac{\hbar}{e}
\frac{q_\alpha J^{\underline{C}C}_{\alpha} ({\bf k}) }{
E_{\underline{C}C} ({\bf k}_+,{\bf k})}.
\label{eq29}
\end{eqnarray}
%
%
According to Fig.~9, the photon-phonon coupling consists of two
contributions, the direct coupling and the indirect, electron-mediated coupling.
	In the commensurate case in which the Hopfield-like processes \cite{Mahan} 
of Fig.~8(c) are neglected, the linear direct coupling between 
external electromagnetic  
fields ($ E_x (-{\bf q})$ is the electric component of the field polarized
along the highly conducting direction)
and  two ${\bf q} \approx 0$ phonon modes under consideration (Fig.~9(a)) is
given by the ground-state average of the coupling Hamiltonian
\begin{eqnarray}
H^{\rm ext}_{1 } &=&
- \sum_{\lambda {\bf q} } E_x (-{\bf q}) \sum_{LL' {\bf k} \sigma} 
\delta P^{L'L}_{x\lambda} ({\bf k},{\bf q}) 
c^\dagger_{L'{\bf k} \sigma} c_{L{\bf k} \sigma}
\nonumber \\
\delta P^{L'L}_{x\lambda} ({\bf k},{\bf q})  &=&
\frac{s_\lambda e}{\sqrt{2N}} 
g^{L'L}_{\underline{\lambda}} ({\bf k}, {\bf k})
\tilde{u}_{x\lambda {\bf q}}
\label{eq30}
\end{eqnarray}
%
%
($s_A = -1$, $s_\varphi = 1$ and the indices 
in the vertex $g^{L'L}_{\underline{\lambda}} ({\bf k}, {\bf k})$ are 
$\underline{A} = \varphi$,
$\underline{\varphi}=A$).
	Here $\tilde{u}_{x\lambda {\bf q}}$ is the amplitude of the electron
oscillations and $\delta P^{L'L}_{x\lambda} ({\bf k},{\bf q})$ is the related
dipole vertex.
	For the incommensurate case, the direct coupling is given 
in a similar way in terms of the current vertex 
\begin{eqnarray}
\delta J^{L'L}_{x\lambda} ({\bf k},{\bf q})  &=&
\frac{s_\lambda e}{\sqrt{2N}} 
g^{L'L}_{\underline{\lambda}} ({\bf k}, {\bf k})
\frac{\partial}{\partial t} \tilde{u}_{x\lambda {\bf q}}.
\label{eq31}
\end{eqnarray}
%
%
	Since the phase phonon field can be written as
$u_{x \varphi {\bf q}} = 2{\rm i} \langle u_A \rangle \varphi_{\bf q}$ \cite{LRA},
we obtain again that the current vertex is proportional to the time derivative 
$\dot{\varphi}_{\bf q}$, and the related equation of motion is analogous to
Eq.~(\ref{eq19}), with $e$, $m$, $n_{xx}^{\rm eff,0}$ and $\omega_0 ({\bf q})$ 
replaced by $\tilde{e}$, $M$, $1/V_0$ and $\omega_{\varphi {\bf q}}$. 
	These parameters enter into the optical conductivity expression through
two adjustable parameters  ($m^*$ and $\omega_{\varphi {\bf q}}$ 
in Eq.~(\ref{eq6}), for example).
 
 
As a result, the direct photon-phonon coupling constant amounts to the replacement of
$c^\dagger_{L'{\bf k} \sigma} c_{L{\bf k} \sigma}$ in Eq.~(\ref{eq30}) by 
$\langle c^\dagger_{L'{\bf k} \sigma} c_{L{\bf k} \sigma} \rangle 
= \delta_{L,C} \delta_{L',C} f_C ({\bf k})$. 
	This result shows that  the coupling function of the 
(Raman-active) mode $u_{A {\bf q}}$, $g^{CC}_{\varphi} ({\bf k}, {\bf k})$, 
vanishes, while the coupling function of the phase phonon mode 
$u_{\varphi {\bf q}}$ is finite and, as mentioned above, is proportional 
to the magnitude of the charge-density wave $q_c$, which is given by 
\begin{equation}
q_c = -e \frac{1}{N} \sum_{{\bf k} \sigma} \sin \varphi ({\bf k}) 
f_C ({\bf k}).
\label{eq32}
\end{equation}
%
%
	Notice that the fact that the phonons ${\bf u}_{\lambda {\bf q}}$ in
Eq.~(\ref{eq30}) couple to external electromagnetic fields through the function
$g^{CC}_{\underline{\lambda}} ({\bf k}, {\bf k})$, rather than 
$g^{CC}_{\lambda} ({\bf k}, {\bf k})$, is a consequence of the fact 
that the variations $V_{\lambda}({\bf r}_i)$ in Eq.~(\ref{eq2}) are 
proportional to the derivative of ${\bf u} _{\lambda i}$ with respect 
to ${\bf r}_i$, rather than to the displacement vector ${\bf u} _{\lambda i}$.

In the Hopfield-like processes the photon momentum is shared 
between an electron-hole pair and a infrared-active phonon in the way described
by the dipole density operator
$$
\sum_{LL' {\bf q}' {\bf k} \sigma} 
\delta P^{L'L}_{x \varphi} ({\bf k}+{\bf q}'-{\bf q},{\bf k},{\bf q}') 
c^\dagger_{L'{\bf k}+{\bf q}'-{\bf q} \sigma} c_{L{\bf k} \sigma}
$$
%
%
in Eq.~(\ref{eq30}), with $g^{L'L}_{A} ({\bf k}, {\bf k})$ 
in the dipole vertex replaced by
$g^{L'L}_{A} ({\bf k}+{\bf q}'-{\bf q}, {\bf k})$.
	The sum over states in the phase phonon branch (${\bf q}'$) 
is restricted to the region where the 
the vertex $g^{L'L}_{A} ({\bf k}+{\bf q}'-{\bf q}, {\bf k})$ of 
Eq.~(\ref{eqA5}) is non-zero.

The indirect, electron-mediated photon-phonon coupling shown in Fig.~9(b), 
on the other hand, is given by  the interband correlation function 
$\chi^{\rm inter}_{1, \lambda} ({\bf q}, \omega)$  defined in  
Appendix, or by Eq.~(4.3) in Ref.~\cite{Rice}.
	According to Eqs.~(\ref{eq28}) and (\ref{eq29}), 
the product of vertices $q^{\underline{C} C}({\bf k},{\bf k}_+)$ 
and $g_{\lambda}^{C \underline{C}}({\bf k}_+,{\bf k})$ 
in this correlation function is proportional to 
$\exp \{ - \phi ({\bf k}) \} J^{\underline{C}C}_{x} ({\bf k})$,
i.e., it is  proportional to the bare electron group velocity
$v^{0}_{x} ({\bf k})$.
	$v^{0}_{x} ({\bf k})$  is an odd function of $k_x$ and, 
consequently,   the coupling function 
$\chi^{\rm inter} _{1, \lambda} ({\bf q}, \omega)$ vanishes 
for both phonon branches.
	The background of this result is the fact that the
electron-phonon coupling $g$ is independent of the electron wave vectors.
	A more complicated form of the infrared selection rules 
and the photon-phase phonon coupling functions can be expected
in the models in which the electron-phonon coupling is related to the 
variations of the bond energies \cite{Schulz} 
(namely, $g$ depends now on the electron wave vectors, introducing 
a new channel in the electron-photon coupling).

Finally, it is important to notice that the common textbook analyses
\cite{LRA,Rice,Gruner} explain the photon-phase phonon coupling  
in the way (see Eqs.~(4.3) and (4.7) in Ref.~\cite{Rice}) 
which is formally equivalent to the replacement of the bare electron 
group velocity in $\chi^{\rm inter} _{1, \lambda} ({\bf q}, {\rm i} \nu_n)$ 
by $v^{0}_{x} (k_{\rm F})$.
	(That is, both the single-particle interband conductivity and 
the photon-phase phonon coupling are given in terms of the dimensionless function
$f(\omega)$ defined in Ref.~\cite{LRA}.)
	Such correlation function is finite indeed and is of the same order as
the expression (\ref{eq32}).
	However, when analyzing complicated CDW systems,
this approximation has to be taken with reservation.

\section{Conclusion}

In this article, we have derived the infrared selection rules for 
the single-particle and collective contributions to the optical conductivity 
in the tetramerized CDW system with the electron-phonon
coupling related to the variations of electron site energies.
	We have shown that the relations between the effective numbers 
of charge carriers and the activation energy of transport coefficients, 
or the MIR optical threshold energy, observed in the
ordinary Peierls systems are naturally explained in
terms of the CDW coherence effects.
	The same coherence effects are expected to be important in the 
correlated multiband systems  with the CDW instability, as well.
	The quantitative analysis of the latter systems is left for future
investigations.

\section*{Acknowledgement}
This research was supported by the Croatian Ministry of Science 
and Technology under Project 119-1191458-0512.

\appendix

\section{Vertex functions and correlation functions in 
the tetramerized Q1D electronic system}

In order to fix the notation, here we rewrite the textbook expressions
\cite{LRA,Schulz,Rice,Gruner}
for the exactly solvable tetramerized CDW case in which the electron-phonon 
coupling is independent of the electron wave vectors 
(Eq.~(\ref{eq2}) in the main text).

The total Hamiltonian reads as 
$H=H_0+H_1'$, with the bare Hamiltonian 
and the electron-phonon coupling Hamiltonian  given, respectively, by
\begin{eqnarray}
H_0 &=&  
\sum_{{\bf k}\sigma}
\varepsilon ({\bf k}) c^{\dagger}_{{\bf k}\sigma} c_{{\bf k}\sigma}   
+ \frac{1}{2M} \sum_{l=\pm 1}\sum_{{\bf q} \approx 0}  
\bigg[p_{{\bf q}+l{\bf Q}}^\dagger p_{{\bf q}+l{\bf Q}}
\nonumber \\ 
&& + \big( M \omega^0_{{\bf q}+l{\bf Q}} \big)^2 
u_{{\bf q}+l{\bf Q}}^\dagger u_{{\bf q}+l{\bf Q}}\bigg], 
 \nonumber \\ 
H_1' &=&  
\sum_{{\bf q}\approx 0 {\bf k} \sigma} \bigg[
\frac{g {\rm e}^{{\rm i} \phi} }{\sqrt{N}} 
\big( u_{A {\bf q}} + {\rm i} u_{\varphi {\bf q}} \big)
c^{\dagger}_{{\bf k}+ {\bf q}\sigma} c_{{\bf k}+ {\bf Q}\sigma}
+ {\rm H.c.} \bigg].
\nonumber \\ 
\label{eqA1}
\end{eqnarray}
%
%
	Here $\varepsilon ({\bf k}) = 
-\sum_{\alpha} 2t_{\alpha} \cos k_\alpha a_\alpha$ is the bare electron
dispersion.
	$p_{{\bf q}+l{\bf Q}}$ is the phonon field conjugate to 
$u_{{\bf q}+l{\bf Q}}$ and $\omega^0_{{\bf q}+l{\bf Q}}$ 
is the bare phonon frequency.

The result of the  phase transformation of $H$, which leads to the replacement 
$u_{A {\bf q}} \rightarrow \delta_{{\bf q},0} \langle u_{A} \rangle 
+ u_{A {\bf q}}$,  is the effective tetramerized  single-electron Hamiltonian
\begin{eqnarray}
\widetilde{H}_0^{\rm e} &=&  
\sum_{ll'{\bf k}\sigma}
H^{ll'}_0 ({\bf k}) c^{\dagger}_{l{\bf k}\sigma} c_{l'{\bf k}\sigma},
\label{eqA2}
\end{eqnarray}
%
%
with the index $l = -1, 0, 1, 2$, and with $\sum_{\bf k}$ running over the 
first Brillouin zone of the tetramerized lattice.  
The matrix elements in the Hamiltonian (\ref{eqA2}) are 
$H^{ll}_0 ({\bf k}) = \varepsilon_l ({\bf k}) 
= \varepsilon ({\bf k}+l{\bf Q})$, representing the dispersions of four
artificially tetramerized bands,
and $H^{ll+1}_0 ({\bf k}) = \Delta \exp \{{\rm i} \phi \}$,
with $\Delta = g \langle u_{A} \rangle /\sqrt{N}$ being the interband
hybridization term.
	The diagonalization of (\ref{eqA2}) leads to 
\begin{eqnarray}
\widetilde{H}_0^{\rm e} &=&  
\sum_{L {\bf k}\sigma}
E_L ({\bf k}) c^{\dagger}_{L{\bf k}\sigma} c_{L{\bf k}\sigma}.
\label{eqA3}
\end{eqnarray}
%
%
	The $E_L ({\bf k})$  are four Bloch energies of the present 
problem (Eqs.~(\ref{eq4}) in the main text), and they,
 together with the total energy of the
electronic system, depend on the phase $\phi$ in $H^{ll+1}_0 ({\bf k})$.

The observation (see the discussion of Fig.~1 in Sec.~2.1) that, 
for $\langle u_{A} \rangle \neq 0$,
the lattice vibrations associated with  
$u_{A {\bf q}}$ and  $u_{\varphi {\bf q}}$ are Raman and infrared
active, respectively, means that there will be the direct photon-phonon 
coupling similar to the photon-phonon coupling in ordinary semiconductors.
	In principle, one can also expect the electron-mediated 
photon-phonon coupling, and, as discussed in detail in Sec.~5, 
in the common microscopic approaches, 
exactly this coupling is regarded as the dominant coupling
\cite{LRA,Schulz,Rice,Gruner}.  
	It is given in terms of the dimensionless 
electron-phonon coupling constants, \\
$g_{\lambda}^{L'L} ({\bf k}_+ ,{\bf k})$ in
\begin{equation}
\widetilde{H}_1' =  
\sum_{\lambda {\bf q}\approx 0 } u_{\lambda {\bf q}}
\sum_{LL'{\bf k} \sigma} \frac{g}{\sqrt{N}}
g_{\lambda}^{L'L} ({\bf k}_+,{\bf k}) 
c^{\dagger}_{L'{\bf k}+ {\bf q}\sigma} c_{L{\bf k}\sigma},
\label{eqA4}
\end{equation}
%
%
and the electron-photon coupling constants (current vertices).

The  coupling constant $g_{\lambda}^{L'L} ({\bf k}_+ ,{\bf k})$ 
in Eq.~(\ref{eqA4})  is given in the usual way
\begin{eqnarray}
g_{\lambda}^{L'L} ({\bf k}_+ ,{\bf k}) &=&  
\sum_l  \big[ {\rm e}^{{\rm i} \phi_\lambda} 
U_{{\bf k}+ {\bf q}}(l,L') U^*_{{\bf k}}(l+1,L) 
\nonumber \\
&& + {\rm e}^{-{\rm i} \phi_\lambda} 
U_{{\bf k}+ {\bf q}}(l+1,L') U^*_{{\bf k}}(l,L) \big]
\label{eqA5}
\end{eqnarray}
%
%
(${\bf k}_+ = {\bf k} + {\bf q}$).
	Here $\phi_A = \phi$ and $\phi_\varphi = \phi + \pi/2$ are two 
useful abbreviations.
	The $U_{{\bf k}}(l,L)$ are the transformation-matrix 
elements defined by
\begin{eqnarray} 
c_{{\bf k} +l{\bf Q} \sigma}^{\dagger} &\equiv& c_{l{\bf k} \sigma}^{\dagger} 
= \sum_{L} U_{\bf k} (l,L)  c_{L{\bf k} \sigma}^{\dagger}.
\label{eqA6}
\end{eqnarray} 
%
%
	The coupling of electrons to  external long-wavelength 
scalar fields  has the form which is similar to Eq.~(\ref{eqA4}),
\begin{equation}
H^{\rm ext} =  
\sum_{{\bf q} } V^{\rm ext} (- {\bf q})
\sum_{LL' {\bf k} \sigma} eq^{LL'} ({\bf k}_+,{\bf k}) 
c^{\dagger}_{L'{\bf k}+ {\bf q}\sigma} c_{L{\bf k}\sigma}.
\label{eqA7}
\end{equation}
%
%
	Here the $q^{L'L} ({\bf k}_+,{\bf k})$  are
the dimensionless monopole-charge vertices, in the intraband channel equal to
unity, and in the interband channel proportional to the interband current 
vertex $J^{L'L}_{\alpha} ({\bf k})$,
\begin{eqnarray}
q^{L'L} ({\bf k}_+,{\bf k}) &=&  
\sum_l U_{{\bf k}+ {\bf q}}(l,L') U^*_{{\bf k}}(l,L)
\label{eqA8}
\\ \nonumber 
&\approx& \delta_{L,L'} + (1 -\delta_{L,L'})
\sum_{\alpha} \frac{(\hbar/e)q_\alpha J^{L'L}_{\alpha} ({\bf k}) }{
E_{L'} ({\bf k}_+) - E_{L} ({\bf k})},
\end{eqnarray}
%
%
with ${\bf q} = \sum_\alpha q_\alpha \hat{e}_\alpha$.	

Finally,  after neglecting the relaxation processes 
in the electron-hole propagators associated with the disorder in the system, 
the renormalization of the phonon 
frequencies in (\ref{eqA1}) is given in terms of the phonon self-energy
\begin{eqnarray}
\hbar \Pi_{\lambda} ({\bf q}, {\rm i} \nu_n) &=& 
g^2 V_0 \, \chi_{\lambda, \lambda} ({\bf q}, {\rm i} \nu_n),
\label{eqA9}
\end{eqnarray}
%
where the correlation function 
$\chi_{\lambda, \lambda} ({\bf q}, {\rm i} \nu_n)$ has the form
\begin{eqnarray}
\chi_{\lambda ,\lambda} ({\bf q}, {\rm i} \nu_n) &=&   
 \frac{1}{V}  \sum_{LL'{\bf k}\sigma}
\big| g_{\lambda}^{L'L}({\bf k}_+,{\bf k}) \big|^2
\nonumber \\
&& 
\times 
\frac{f_{L'}({\bf k}_+)-f_{L}({\bf k})}{{\rm i} \hbar \nu_n
+E_{L'}({\bf k}_+)-E_{L}({\bf k})}.
\label{eqA10}
\end{eqnarray}
%
%
	In the same approximation, the electron-mediated coupling of 
phonons to the external scalar fields is given by 
$\chi_{\lambda, 1} ({\bf q}, {\rm i} \nu_n)$.
	This correlation function is obtained by replacing the 
vertices $\big| g_{\lambda}^{L'L}({\bf k}_+,{\bf k}) \big|^2$ in
Eq.~(\ref{eqA10})
by $g_{\lambda}^{L'L}({\bf k}_+,{\bf k}) $ $ q^{LL'}({\bf k},{\bf k}_+)$. 
	Similarly, the charge-charge correlation function 
$\chi_{1, 1} ({\bf q}, {\rm i} \nu_n)$ is given by Eq.~(\ref{eqA10})
with $|q^{LL'}({\bf k},{\bf k}_+)|^2$ being the vertices in question.
	For all three correlation functions we obtain 
$\chi_{i,j} ({\bf q}, \omega)$, $i,j \in \{ 1, \lambda\}$, by the 
analytical continuation of $\chi_{i,j} ({\bf q},  {\rm i} \nu_n)$.

\end{document}